\documentclass{article}
\usepackage{spconf,amsmath,graphicx,bm, booktabs, verbatim,tabularx}
\newcolumntype{C}[1]{>{\centering\arraybackslash}p{#1}}
\title{Investigation of Monaural Front-End Processing for Robust ASR without Retraining or Joint-Training}

\name{$^1$Zhihao Du, $^2$Xueliang Zhang, $^1$Jiqing Han\thanks{The work was performed while the first author was a trainee at Elevoc Technology Co.,Ltd., ShenZhen, China.}}
\address{$^1$School of Computer Science and Technology, Harbin Institute of Technology, Harbin, China\\
	$^2$Department of Computer Science,	Inner Mongolia University, Hohhot, China\\
	\texttt{zhihaodu.china@gmail.com, cszxl@imu.edu.cn, jqhan@hit.edu.cn}}

\begin{document}

\maketitle

\begin{abstract}
	
In recent years, monaural speech separation has been formulated as a supervised learning problem, which has been systematically researched and shown the dramatical improvement of speech intelligibility and quality for human listeners.
However, it has not been well investigated whether the methods can be employed as the front-end processing and directly improve the performance of a machine listener, i.e., an automatic speech recognizer, without retraining or joint-training the acoustic model.
In this paper, we explore the effectiveness of the independent front-end processing for the multi-conditional trained ASR on the CHiME-3 challenge. We find that directly feeding the enhanced features to ASR can make 36.40\% and 11.78\% relative WER reduction for the GMM-based and DNN-based ASR respectively. We also investigate the affect of noisy phase and generalization ability under unmatched noise condition. 
	
\end{abstract}


\section{Introduction}

Monaural speech separation aims at separating speech from the noisy backgrounds by using one microphone. In recent years, speech separation has been formulated as a supervised learning problem. Thanks to the rise of deep learning, supervised speech separation has made significant progress \cite{wang2018supervised}. 

Speech separation for improving human speech intelligibility and quality has been systematically evaluated and successfully utilized. 
In general, speech separation can be divided into three groups, i.e., masking-based methods, mapping-based methods and signal approximation.
The masking-based methods try to predict a mask computed from premixed noise and clean speech, e.g. ideal ratio mask \cite{wang2014training}, phase sensitive mask \cite{erdogan2015phase} and complex ratio mask \cite{williamson2016complex}. Mapping-based method tries to enhance speech by finding a mapping function between noisy feature and spectrum of the clean speech \cite{xu2015regression}.
The idea of signal approximation (SA) is to train a ratio mask estimator that minimizes the difference between the spectral magnitude of clean speech and that of estimated speech \cite{weninger2014discriminatively}.
A lot of learning machines have also been introduced for speech separation, In \cite{wang2014training, williamson2016complex}, deep neural networks (DNNs) are employed to predict ideal masks. Lu \emph{et al.} used a deep denoising auto-encoder (DDAE) to obtain a clean Mel frequency power spectrogram (fbank) from a noisy one \cite{lu2013speech}, In \cite{hui2015convolutional, tan2018gated}, convolutional neural networks (CNNs) have been introduced. Besides the feed-forward networks, recurrent networks (RNNs) have also became a popular choice in the speech separation community \cite{erdogan2015phase}. 
As for features, Wang \emph{et al.} proposed a complementary feature \cite{wang2013exploring} and Chen \emph{et al.} found multi-resolution cochleagram is a better feature in low signal-noise-ratio conditions \cite{6854965}.

Compared with human listeners, ASR is more sensitive to the noise interfering and the speech distortion. In general, there are three strategies introduced to improve the robustness of ASR.
The first one is using a separation frontend to enhance both training and test sets and retraining the acoustic model with enhanced features \cite{han2015deep, weninger2015speech}. 
The second one is joint-training the front-end enhancement model with the back-end acoustic model \cite{wang2016joint,liu2018boosting}. 
The third one is multi-conditional training which performs acoustic modeling on noisy speech and the extracted features are directly fed to the acoustic model for decoding at the test stage. This strategy is shown to be effective in matched condition but gives an unremarkable performance for the unseen noise \cite{li2014long}.

All the above strategies require retraining or joint-training an acoustic model which can be time-consuming and sophisticated. 
Compared with speech separation, it is relatively hard to collect training data for speech recognition which needs handcrafted annotation.
In practice, a preferred choice is to train the front-end speech separation and the back-end ASR independently. And we wonder whether the supervised speech separation methods can directly improve the performance of ASR without retraining or joint-training under the real noisy condition. 
Wang \emph{et al.} evaluated a masking-based method on the simulated noisy dataset which is derived from Google Voice dataset, which made 0.3\% improvement for the multi-condition trained ASR \cite{wang2015time}. 
Wang \emph{et al.} investigated the effectiveness of the front-end processing on the reverberant condition \cite{Wang2018Inves}. 
But there still lacks of a work to systematically examine the ability of the supervised speech separation methods for the multi-conditional trained ASR. In this paper, different speech separation methods based on various time-frequency (T-F) representations are investigated on the third CHiME challenge. 

\section{Speech separation methods}

In speech separation community, RNNs with long short-term memorys (LSTMs) have been widely employed to leverage the sequential information of speech signals and shown superior performance as compared with DNNs and CNNs \cite{erdogan2015phase, Wang2018Inves}.
For optimization objectives, ratio masking, direct mapping and signal approximation are three popular choices. Note that all these methods can be performed in different T-F representations, such as log-power spectrogram, log-fbank feature.
In this investigation, we wonder which combination of the optimization objectives and T-F representations is most appropriate for the robust ASR. Therefore, we fix our learning machine as a RNN with the bidirectional long short-term memories (BiLSTMs) \cite{schuster1997bidirectional} and focus on the different optimization objectives and T-F representations.

\subsection{Optimization objectives}
The general training objective of supervised speech separation is defined as:
\begin{equation}
	\min_\Phi\sum_{n=1}^{N}{\ell(\bm{d}_n, f_\Phi(\bm{y}_n))}
\end{equation}
where $\bm{d}_n$ is the desired output at frame $n$, $\bm{y}_n$ is the noisy T-F representation and the input of separation model $f_\Phi(\cdot)$ which is parameterized by $\Phi$, and $\ell(\cdot, \cdot)$ means \emph{squared loss}, which is defined as:
\begin{equation}
\ell(\bm{x},\bm{y}) = {\lVert \bm{x}-\bm{y} \rVert}_2^2
\end{equation}
where $\lVert\cdot\rVert$ is the 2-norm of a vector.
\subsubsection{Ratio Masking}
The masking-based methods try to learn a mapping function from the noisy T-F representations to the T-F masks of the clean speech. The training target of the ratio mask is defined as:
\begin{equation}
	\min_\Phi\sum_{n=1}^{N}{\ell(\bm{m}_n, f_\Phi(\bm{y}_n))}
\end{equation}
where $\bm{m}_n$ is the desired ratio mask at frame $n$. We investigate a direct masking method, which is defined as:
\begin{equation}
	\bm{m}_n = \frac{\bm{s}_n}{\bm{y}_n}
\end{equation}
where $\bm{s}_n$ and $\bm{y}_n$ are the T-F representations of clean and noisy speech at frame $n$ respectively. Because the direct masks are not well bounded, we clip them to $[0, 1]$ for the training stability.
\subsubsection{Direct Mapping}
Mapping-based methods train the learning machine to predict the T-F representation of the clean speech from the noisy speech directly. The optimization objective of direct mapping is defined as:
\begin{equation}
	\min_\Phi\sum_{n=1}^{N}{\ell(\bm{s}_n, f_\Phi(\bm{y}_n))}
\end{equation}
where $\bm{s}_n$ and $\bm{y}_n$ are the T-F representations of the clean and noisy speech at frame $n$ respectively.
\subsubsection{Signal Approximation}
SA-based methods implicitly learn ratio mask from noisy T-F representations. Different from the masking-based methods which directly reduce the training loss between the desired mask and the predicted one, SA-based methods reduce the loss between the T-F representations of the target speech and the estimated ones. SA-based optimization objective is defined as:
\begin{equation}
	\min_\Phi\sum_{n=1}^{N}{\ell(\bm{s}_n, \bm{y}_n \odot f_\Phi(\bm{y}_n))}
\end{equation}
where $\odot$ is element-wise multiplication. The output of $f_\Phi(\bm{y}_n)$ is restricted to the range $[0,1]$ and bounded as the ratio mask.
\subsection{Target domains}
The above optimization objectives can be performed on different target domains. In ASR community, log-fbank is the most used feature, so we optimize our model on the log-fbank domain. Because the log-fbank features can be directly extracted from the spectrograms (fft domain), we also perform the optimization on the fft domain and its logarithmic counterpart. 
\subsection{Features}
Different learning tasks can benefit from the appropriate features. Log-fbank features are widely used for training the acoustic model, and the log-fft spectrograms are usually fed to the speech separation models. In this paper, the targets on log-fbank and fbank domain are predicted from log-fbank features and the log-fft spectrograms are fed to the models on log-fft and fft domain.
The input features, output domains and optimization objectives of the evaluated methods are shown in table \ref{tab:input_output}.
\linespread{0.8}
\begin{table}[!thb]
	\centering
	\caption{The input features, output domains and optimization objectives of the evaluated methods.}
	\setlength{\tabcolsep}{1.2mm}
	\begin{tabular}{p{2.7cm}|p{1.4cm}p{1.4cm}p{2cm}}
		\toprule
		Evaluated\ \ \ \ \ \ \ \ \ \ methods & Input\ \ \ \ domain & Output domain & Optimization objectives \\
		\midrule
		log-fbank mapping & log-fbank & log-fbank & mapping\\
		log-fbank SA & log-fbank & log-fbank & SA\\
		log-fbank masking & log-fbank & log-fbank & ratio masking\\
		\midrule
		log-fft mapping & log-fft & log-fft & mapping\\
		log-fft SA & log-fft & log-fft & SA\\
		log-fft masking & log-fft & log-fft & ratio masking\\
		\midrule
		fbank masking & log-fbank & fbank & ratio masking\\
		fft masking & log-fft & fft & ratio masking\\
		\bottomrule
	\end{tabular}
	\label{tab:input_output}
\end{table}

\section{Experimental settings}

We perform our investigation on the CHiME-3 Challenge \cite{barker2015third} which provides multi-channel data for distant-talking automatic speech recognition and we only use the fifth channel in this paper. 

In the training phase of ASR, we follow the recipe for CHiME-3 in the newest Kaldi release to build our baseline. 
There are two differences between our training and the default. First, we train the recognizer with multi-conditional training strategy (MCT), i.e., we train the GMM-based and DNN-based acoustic model with the clean utterances, the simulated noisy utterances in the fifth channel, the real noisy utterances in the fifth channel and the real close-talk utterances in channel zero while the default training is only with the real and simulated noisy utterances in the fifth channel. The intuition behind this MCT is that the front-end processing tries to reconstruct the clean features, only training the recognizer with the noisy utterances is obviously unreasonable. 
Second, we train the recognizer with fbank features instead of MFCC. The fbank feature has been widely used in robust speech recognition community \cite{li2014overview}. With the MCT strategy and fbank features, our ASR baseline achieves the similar performance which is claimed in the CHiME-3 challenge.

For the front-end processing, we employ a 4-layer RNN with 512 bidirectional LSTM cells in each layer. A dense layer with softplus activations is followed for the mapping-based methods. And the sigmoid function is employed for the masking-based and the SA-based methods. Different methods are evaluated on the log-fft domain and log-fbank domain, however, the fft and fbank domain are only evaluated with the masking-based method because of their large value range. To evaluate the affect of noisy phases, the recognizer is also fed with the synthesized waveforms which are reconstructed from the noisy phases and the estimated magnitudes via the inverse STFT. 
In the training phase of the front-end models, the T-F representations extracted from the simulated and real noisy utterances are fed to the models and the corresponding clean counterparts are estimated. 
We also expand the training set by mixing the clean utterances and the noise records in training set by 0dB, 3dB and 6dB.

In evaluating phase, the word error rate (WER) is calculated for the simulated and real noisy utterances in development and test set. The front-end processing is also performed on the clean and close-talk utterances to find whether it will lead to a degradation on the relatively clean utterances.

\linespread{0.36}
\begin{table*}[!htb]
	\centering
	\caption{The WERs (\%) of GMM-based ASR on development and test set}
	\setlength{\tabcolsep}{2.1mm}
	\begin{tabular}{l|cccc|cccc|cc}
		\toprule
		Methods & dt\_bth & dt\_close & dt\_simu & dt\_real & et\_bth & et\_close & et\_simu & et\_real & dt\_avg & et\_avg \\ 
		\midrule
		Baseline & 5.63 & 7.52 & 20.26 & 21.29 & 5.60 & 14.31 & 25.00 & 38.39 &	20.78 &	31.70 \\
		\midrule
		log-fbank mapping & 6.31 &	7.60 &	16.87 &	16.48 &	6.39 &	11.05 &	18.40 &	28.56 & 	16.68 &	23.48 \\
		log-fbank SA      & 5.68 &	6.98 &	14.99	&	\textbf{15.28} &	5.81 &	9.99 &	16.88 &	25.87 & \textbf{15.14} &	21.38 \\
		log-fbank masking & 5.74 &	7.15 &	15.15 &	15.54 &	5.85 &	10.04 &	16.98 &	25.65 & 15.35 &	21.32 \\
		\midrule
		log-fft mapping & 6.31 & 8.25 &	18.99 &	19.71 &	6.13 &	12.14 &	22.22 &	30.26 &	19.35 &	26.24 \\
		\ \ \ \ +noisy phases & 6.42 &	8.13 &	18.00 &	19.76 &	6.52 &	12.20 &	20.73 &	30.34 &	18.88 &	25.54 \\
		log-fft SA      & 5.93 &	7.37 &	17.40 &	17.87 &	6.01 &	11.07 &	19.83 &	28.18 &	17.64 &	24.01 \\
		\ \ \ \ +noisy phases           & 5.94 &	7.30 &	16.56 &	17.56 &	5.85 &	11.04 &	18.77 &	27.88 &	17.06 &	23.33 \\
		log-fft masking & 5.78 &	7.44 &	16.66 &	17.54 &	5.85 &	11.56 &	19.54 &	27.94 &	17.10 &	23.74 \\
		\ \ \ \ +noisy phases & 6.11 &	7.30 &	16.27 &	16.92 &	5.66 &	11.45 &	18.69 &	27.85 &	16.60 &	23.27 \\
		\midrule
		fbank masking   & 5.69 &	7.15 &	\textbf{14.19} &	17.01 &	5.88 &	9.66 &	\textbf{15.36} &	24.95 & 15.60 & \textbf{20.16} \\
		fft masking     & 5.56 &	7.09 &	14.48 &	16.19 &	5.73 &	10.22 &	16.16 &	\textbf{24.84} & 15.34 & 20.50 \\ 
		\ \ \ \ +noisy phases & 5.99 &	7.26 &	14.51 &	16.39 &	5.77 &	14.38 &	17.06 &	27.67 &	15.45 &	22.37 \\
		\bottomrule
	\end{tabular}
	\label{tab:wer_gmm}
\end{table*}
\linespread{0.36}
\begin{table*}[!htb]
	\centering
	\caption{The WERs (\%) of DNN-based ASR on development and test set}
	\setlength{\tabcolsep}{2.1mm}
	\begin{tabular}{l|cccc|cccc|cc}
		\toprule
		Methods & dt\_bth & dt\_close & dt\_simu & dt\_real & et\_bth & et\_close & et\_simu & et\_real & dt\_avg & et\_avg \\
		\midrule
		Baseline          & 3.42 &	4.92 &	12.68 &	14.19 &	4.03 &	8.11 &	15.14 &	25.44 & 13.44 &	20.29  \\
		\midrule
		log-fbank mapping & 4.04 &	5.18 &	13.64 &	14.40 &	4.58 &	8.36 &	15.10 &	26.34  & 14.02 & 20.72\\
		log-fbank SA      & 3.36 &	4.77 &	12.30 &	12.95 &	4.09 &	7.13 &	13.78 &	22.73  & 12.63 & 18.26\\
		log-fbank masking & 3.36 &	4.73 &	\textbf{12.08} &	\textbf{12.70} &	3.92 &	7.12 &	13.44 &	22.36 & \textbf{12.39} &	\textbf{17.90} \\
		\midrule
		log-fft mapping   & 3.92 &	5.94 &	16.57 &	16.14 &	4.52 &	9.74 &	19.38 &	25.87  & 16.36 &	22.63\\
		\ \ \ \ +noisy phases & 3.98 &	5.93 &	16.49 &	16.09 &	4.76 &	10.07 &	19.35 &	25.90 &	16.29 &	22.63 \\
		log-fft SA        & 3.47 &	5.04 &	14.84 &	14.72 &	4.11 &	8.03 &	17.25 &	24.78  & 14.78 &	21.02\\
		\ \ \ \ +noisy phases & 3.89 &	5.08 &	14.66 &	14.34 &	4.20 &	8.17 &	16.81 &	24.20 &	14.50 &	20.51 \\
		log-fft masking   & 3.50 &	5.13 &	14.59 &	14.17 &	4.09 &	8.73 &	17.26 &	24.89 &	14.38 &	21.08\\
		\ \ \ \ +noisy phases & 3.69 &	5.18 &	14.49 &	13.97 &	4.30 &	8.93 &	17.08 &	24.70 &	14.23 &	20.89\\
		\midrule
		fbank masking     & 3.32 &	4.87 &	12.46 &	14.63 &	4.13 &	6.93 &	\textbf{13.42} &	23.50 & 13.55 &	18.46\\
		fft masking		  & 3.38 &	4.93 &	12.09 &	13.42 &	4.15 &	7.46 &	13.65 &	\textbf{22.26} &  12.76 & 17.96\\
		\ \ \ \ +noisy phases & 3.73 &	5.08 &	12.23 &	13.13 &	4.24 &	10.57 &	14.35 &	24.21 &	13.74 &	19.28 \\
		\bottomrule
	\end{tabular}
	\label{tab:wer_dnn}
\end{table*}
\section{Results and discussions}

Table \ref{tab:wer_gmm} and \ref{tab:wer_dnn} show the WERs of GMM-based and DNN-based ASR respectively. 
The columns with \texttt{dt\_*} and \texttt{et\_*} show the results of development set and test set.
The WERs of utterances recorded in booth and real noisy environments are given in columns \texttt{*\_bth} and \texttt{*\_real}.
The columns \texttt{*\_close} represent the results of close-talk utterances in channel zero and the WERs of simulated noisy speech in fifth channel are shown in columns \texttt{*\_simu}.
The rows marked by "\texttt{+noisy phases}" indicate that we reconstruct waveforms in time domain and extract the ASR features on the waveforms. We do this because that speech enhancement often runs in local system and ASR locates in cloud server for many real scenarios, and the interface always needs waveform.
The average performances of simulated and real noisy utterances are given in the \texttt{*\_avg} columns.

For the GMM-based ASR (seen in table \ref{tab:wer_gmm}), masking-based method in the log-fbank domain achieves the best performance, 36.40\% relative improvement from 31.70\% to 20.16\%, on the noisy test set. SA in the log-fbank domain gets the lowest WER on the noisy development set. It seems that the mapping-based method is not a good ideal for the automatic speech recognition purpose. When noisy phase is involved, the masking-based method in the fft domain degrades significantly on test set. Although the methods in the log-fft domain are affected slightly by noisy phase, performances are much worse than the methods in the fft domain.

For the DNN-based ASR(seen in table \ref{tab:wer_dnn}), the masking-based method in the log-fbank domain is a good choice and achieves 7.78\% and 11.78\% relative improvement on the noisy development and test set respectively. The masking-based method in the fft domain gets lower WER than all methods in the log-fft domain, but it is significantly degraded by noisy phase. The mapping-based front-end processing and the methods in the log-fft domain do not improve the performance of ASR anymore. 

These front-end processing methods make very little degradation on the relatively clean speech utterances (see the \texttt{*\_clean} and \texttt{*\_close} columns). Surprisingly, some methods can even improve the performance of ASR for the close-talk utterances in test set, which is possibly because the close-talk utterances are not very clean but slightly noisy.

From table \ref{tab:wer_dnn} and \ref{tab:gener_ability}, we can see that independent front-end processing can dramatically enhance the ASR performance with same noise condition.
To evaluate the generalization ability, we calculate WERs of noisy utterances interfered by babble noise which does not appear in ASR and speech enhancement training data. 
The method in the log-fbank domain achieves the best performance for the unseen babble noise which also gets the lowest WER under the noise matched condition. We find the ASR with MCT strategy does not generalize well for the unseen noise while speech enhancement efficiently leverages the information of noise and performs better under the unmatched condition.
\section{Conclusions}

In this paper, we investigate the independent front-end processing methods for ASR without retraining or joint-training on the CHiME-3 challenge. The masking-based, mapping-based and SA-based methods are evaluated in the log-fbank domain, log-fft domain and their linear counterparts. From this investigation, we find the masking-based method is a good choice for ASR. 
Direct masking in the log-fbank domain achieves the lowest WER under the matched and unmatched noise condition as compared with the baseline which is a strong DNN-based acoustic model. 
\linespread{1.0}
\begin{table}[!thb]
	\centering
	\caption{The WERs (\%) of the masking-base methods under the unmatched noise condition.}
	\setlength{\tabcolsep}{1.4mm}
	\begin{tabular}{p{2.8cm}|C{1.5cm}C{1.5cm}C{1.5cm}}
		\toprule
		Methods & 0 dB & 3 dB & 6 dB\\
		\midrule
		Baseline & 38.22 &	21.53 &	13.23\\
		\hline
		log-fbank masking & \textbf{32.48} &	\textbf{18.21} &	\textbf{10.54}\\
		fbank masking & 34.89 &	19.29 &	11.09 \\
		log-fft masking & 39.52 &	22.96 &	13.76\\
		fft masking & 34.56 &	19.73 &	11.82 \\
		\bottomrule
	\end{tabular}
	\label{tab:gener_ability}
\end{table}
Noisy phase leads to a considerable degradation for the masking-based methods in the fft domain while the affect in the log-fft domain is very slight. The independent front-end generalizes better than MCT for the unseen noise. In the future, we will try to further reduce WER of the DNN-based ASR with the independent front-end processing.

\section{Acknowledgements}

This research was supported by National Science Foundation of China No.61876214, National Key Research and Development Program of China under Grant 2017YFB1002102 and National Natural Science Foundation of China under Grant U1736210.

\vfill\pagebreak

\label{sec:refs}
\bibliographystyle{IEEEbib}
\bibliography{refs}

\end{document}